%% file: main.tex
\documentclass[pdflatex,sn-apa]{sn-jnl}
\usepackage{amsmath,amssymb,amsfonts}
\usepackage{algorithmic}
\usepackage{graphicx}
\usepackage{booktabs}
\usepackage{multirow}
\usepackage{tabularx}
\usepackage{needspace}
\usepackage[position=bottom]{subfig}
\usepackage{pgfplots}
\pgfplotsset{compat=1.18}
\usepgfplotslibrary{groupplots}
\usetikzlibrary{positioning}
\usepackage{adjustbox}
\usepackage{textcomp}
\usepackage{xcolor}
\usepackage{colortbl}
\usepackage{url}
\usepackage{hyperref}

\IfFileExists{Orcidlogo.eps}{}{%
  \AtBeginDocument{\gdef\orcidlogo{\textsuperscript{\scriptsize\textsc{id}}}}%
}

\def\BibTeX{{\rm B\kern-.05em{\sc i\kern-.025em b}\kern-.08em
    T\kern-.1667em\lower.7ex\hbox{E}\kern-.125emX}}

\begin{document}
\title{Realization Divergence in Multi-Provider Quantum Computing: A Pre-Registered
Study of What Survives Measurement}

\author*[1,2]{\fnm{Jouni} \sur{Peltonen}\orcid{https://orcid.org/0009-0000-9673-9827}}\email{jouni.s.peltonen@jyu.fi}
\author[1]{\fnm{Vlad} \sur{Stirbu}\orcid{https://orcid.org/0000-0001-9462-5922}}\email{vlad.a.stirbu@jyu.fi}
\author[1]{\fnm{Tommi} \sur{Mikkonen}\orcid{https://orcid.org/0000-0002-8540-9918}}\email{tommi.j.mikkonen@jyu.fi}

\affil[1]{\orgname{University of Jyväskylä}, \orgaddress{\city{Jyväskylä}, \country{Finland}}}
\affil[2]{\orgname{QMill Oy}, \orgaddress{\city{Espoo}, \country{Finland}}}

\maketitle

\begin{abstract}

Quantum programs pass through compilers, transpilers and vendor services before
execution. However, the distance between what was written and what runs is rarely measured. We
call this distance \emph{realization divergence}: the observable difference in how an
identical source circuit is realized and behaves across quantum software stacks. We report
a hardware study across three vendors and four devices, combining pre-registered
predictions with explicitly identified exploratory analyses and amendments.
In a compilation experiment with assigned placement policies, embedding radius correlates
with two-qubit overhead at Spearman 0.805, against 0.492 for circuit size.
Under a fixed lowest-index policy, an eight-qubit Bernstein--Vazirani circuit exhibits
greater execution divergence than a twelve-qubit circuit on IQM Emerald. The opposite
ordering occurs on IQM Garnet. Both comparisons replicate across two repetitions with
disjoint confidence intervals. These results support a placement-dependent explanation,
without isolating individual hardware noise mechanisms. The same ordering inversion is
absent on the all-to-all IonQ control. On Rigetti Cepheus, a follow-up using topology-aware
native placement found no consistent advantage in correct-answer probability for either
submission track, but did not establish equivalence between the tracks.
Structural comparisons depend on metric weighting,
placement and device generation. Provider interfaces also limit observability: Rigetti's
supported-operation list remained unchanged across a device generation that replaced the
native entangler. We classify silent interface failures in composed toolchains and identify
checks at toolchain boundaries, connecting valid cross-provider comparisons with concrete
software-quality practices.
\end{abstract}


\section{Introduction}
\label{introduction}

Cloud-based access to multi-provider quantum platforms has enabled quantum algorithms to be routinely developed, executed, and compared across providers~\citep{nguyen2024quantumcloudcomputingreview}. In this environment, quantum software is increasingly expected to be \textit{portable}, \textit{reproducible}, and \textit{comparable} across platforms.

Identical circuit specifications can produce different observed outcomes across
providers. Noise limits current hardware~\citep{Preskill_2018}, while compilation and
qubit mapping affect the circuits exposed to that noise~\citep{li2019tacklingqubitmappingproblem}.
Efficient development workflows therefore need to account for both software and execution
context~\citep{Kinanen2025}.

Quantum software engineering research identifies cross-platform tooling and quality
assurance as continuing challenges~\citep{Murillo_2025}. Existing compiler comparison,
benchmarking and testing approaches provide complementary views of those challenges.
Here we study \emph{realization divergence}: observable differences in the compiled
structure and execution behavior of identical source circuits. Our focus is the validity
of comparisons made from artifacts exposed by heterogeneous provider interfaces, not a
separation of software effects from hardware noise that those artifacts cannot establish.

We use a layered observation model and an integrated toolchain to study these comparisons
through standard SDK artifacts. The contribution is the combined experimental evidence
connecting measurement validity, provider observability and software-quality checks.
It is not a new distance metric or the observation that routing affects performance.
The study tests which comparisons survive changes in placement and measurement choices,
and identifies interface checks needed to interpret them. It spans three vendors and four
devices, combining registered predictions with explicitly identified exploratory analyses.

Following Design Science Research principles~\citep{10.5555/2017212.2017217}, we develop the framework as a practical artifact and use it to measure representative circuits on commercial quantum platforms. The framework is the instrument; what is evaluated is which claims about realization divergence survive being measured with it, what a provider's interface permits measuring at all, and what fails when its components are composed.

This paper makes three contributions:
\begin{itemize}
    \item \textbf{Evidence for conditional realization-divergence orderings.} Experiments
    across several circuit families show why divergence comparisons require explicit
    conditions. Targeted placement experiments and hardware replications support an
    embedding-based explanation of selected divergence orderings. Structural magnitudes
    and rankings depend on weighting,
    placement policy and device generation; execution comparisons also require explicit
    reference distributions and uncertainty estimates.
    \item \textbf{An empirical characterization of provider-interface observability.}
    We identify differences in available execution records and admissible native parameters
    across providers, and show that a stable supported-operation list need not expose
    changes in a device's native gate set.
    \item \textbf{An empirically grounded classification of interface failures in composed
    quantum toolchains.} Observed failures involve needed-but-undeclared,
    declared-but-unconsumed and accepted-but-overridden contracts. These can cause explicit
    errors or produce valid-looking artifacts with inconsistent relationships. We identify
    boundary checks for the observed cases.
\end{itemize}

The rest of this paper is organized as follows. Section~\ref{relatedwork}
introduces the background and related work.
Sections~\ref{methodology} and \ref{framework} present the methodology, research questions
and divergence analysis framework. Section~\ref{demonstration} demonstrates the framework
on commercial quantum hardware, and Section~\ref{evaluation} evaluates the claims it
supports. Section~\ref{sec:discussion} interprets the results, and
Section~\ref{sec:implications} develops their engineering implications.
Sections~\ref{sec:threats} and \ref{sec:future_work} address limitations and future work.
Section~\ref{conclusion} concludes the paper.

\section{Background and Related Work}
\label{sec:backgrnd}
\label{relatedwork}

The quantum computing ecosystem comprises diverse hardware architectures, software toolchains, and execution environments. Cloud platforms such as Amazon Braket~\citep{gonzalez2021braket} and Azure Quantum enable heterogeneous quantum processor access through provider-specific programming models, transpilers, and runtime services~\citep{moguel2022quantum}. Recent surveys highlight the growing heterogeneity in quantum programming interfaces and the challenges of managing cross-provider compatibility~\citep{https://doi.org/10.1049/qtc2.12096}. While accelerating innovation, this diversity introduces substantial complexity for developers and researchers~\citep{10621693}.

In typical workflows, platform-agnostic circuit specifications undergo multiple compilation stages: gate decomposition, qubit mapping, circuit rewriting, and scheduling to produce hardware-executable forms~\citep{Weder2022}. These transformations are essential but largely opaque and vary significantly across providers—even with identical input specifications and controlled transpilation parameters— making the relationship between algorithmic intent and hardware realization difficult to interpret.

Such divergence complicates reproducibility, comparability, and debugging~\citep{10.1145/3643667.3648226}.
A \emph{realization-divergence} perspective follows the observable artifacts through
the software stack: compilation can preserve ideal behavior while changing circuit
structure and the noise exposure of its execution. We distinguish ideal unitary
equivalence, a formal property assessed by equivalence-checking methods~\citep{sander2025equivalencecheckingquantumcircuits},
from structural divergence in compiled circuits and behavioral divergence in measured
output distributions. This paper measures the latter two and studies their relationship;
neither is a test of ideal unitary equivalence.

\subsection{Motivating Example: Bell Circuit Divergence}
\label{sec:bell_motivation}

\begin{figure*}[t]
    \centering
    \footnotesize
    \subfloat[Bell circuit (source)]{\includegraphics[width=0.20\textwidth]{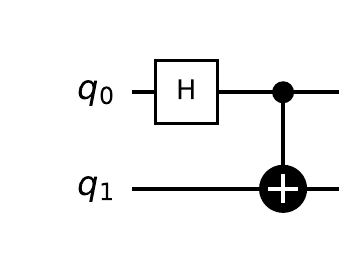}}
    \hfill
    \subfloat[IQM Emerald, native]{\includegraphics[width=0.36\textwidth]{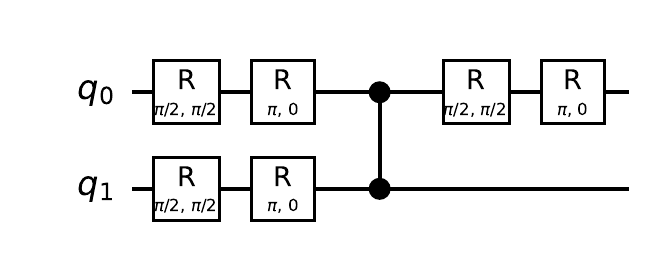}}
    \hfill
    \subfloat[Rigetti Cepheus, native]{\includegraphics[width=0.40\textwidth]{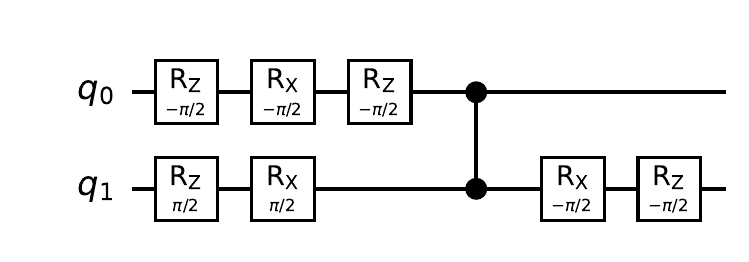}}
    \caption{Two-qubit Bell preparation and its native-gate realizations on two devices.
    Active wires are relabelled locally as $q_0$ and $q_1$ for display; these are not
    device-qubit identifiers. At the supported gate set every evaluated device returns
    the source unchanged. IonQ Forte-1 is omitted because native compilation requires
    a vendor-specific plugin. Panels are generated from stored device profiles by
    \texttt{scripts/build\_circuit\_figures.py} in the research
    artifact~\citep{peltonen2026divartifact}}
    \label{fig:bell_compilation}
\end{figure*}

A two-qubit Bell preparation circuit makes the problem concrete. Its source contains a
Hadamard gate followed by a controlled-NOT gate. Figure~\ref{fig:bell_compilation}
compares that source with two native-gate realizations produced by the client-side compiler.
IQM Emerald requires seven gates; Rigetti Cepheus requires eight, using a different
entangling basis. No routing is needed in this example, so the structural difference
comes from gate-set synthesis rather than connectivity.

A supported gate set lists operations accepted by the provider's interface, whereas a
native gate set describes device-level operations. In this example, compiling to the
supported gate set leaves the source unchanged for every evaluated device, so a comparison
at that layer would miss the native-gate differences shown in Figure~\ref{fig:bell_compilation}.
This example shows why a portability claim must name its observation
layer. Section~\ref{sec:bell_interpretation} revisits the example using the framework;
the subsequent Bernstein--Vazirani (BV) case study demonstrates cross-layer analysis, and
Section~\ref{evaluation} tests the research claims.

\subsection{Quantum Languages and Compilation}

Quantum software development relies on high-level languages and SDKs compiled into
executable circuits~\citep{Weder2022}. Representations such as
OpenQASM~\citep{cross2017openquantumassemblylanguage} describe the input circuits;
mapping and optimization adapt them to a target~\citep{li2019tacklingqubitmappingproblem}.
Retargetable compilation supports multiple native bases and architectures~\citep{Sivarajah_2020}.

Compiler comparison is itself an established research topic. \citet{10.1007/978-3-030-87568-8_4}
automate translation, compilation and comparison across quantum compilers.
MQT Predictor combines device selection with device-specific compilation and optimization
for objectives including expected fidelity~\citep{10.1145/3673241}. These approaches go
beyond syntactic translation and directly address compilation choices.

Reinforcement learning has also been applied to compiler-pass selection~\citep{quetschlich2023compiler}.
Software benchmarking characterizes circuit creation, manipulation and compilation
performance~\citep{nation2025benchmarking,Quetschlich_2023}. Verified optimization provides
correctness guarantees for transformations~\citep{10.1145/3434318}. Our study complements
these approaches by testing which comparisons remain defensible when client-visible
artifacts and hardware records come from different provider interfaces.

\subsection{Benchmarking and Cross-Platform Evaluation}

Benchmarking evaluates quality, speed and scale~\citep{wack2021qualityspeedscalekey},
application performance~\citep{tomesh2022supermarqscalablequantumbenchmark,Lubinski_2023},
and quantum volume~\citep{Cross_2019,9805433}. Compilation is not uniformly treated as
a black box: QUEST, a circuit-approximation method, trades gate count against output
fidelity~\citep{10.1145/3503222.3507739}. This paper examines placement, metric choice,
submission track and observability jointly as conditions on multi-provider comparisons.

\subsection{Quantum Circuit Equivalence and Optimization}
ZX-calculus supports circuit simplification~\citep{Duncan_2020}, and decision-diagram
methods support equivalence checking~\citep{Burgholzer_2021}. QDiff uses differential
testing of quantum software stacks to detect discrepancies~\citep{9678792}.
These methods address correctness questions that structural counts cannot decide.
Our measurements complement those correctness checks by characterizing compiled
structure and noisy execution behavior.

\subsection{Reproducibility, Software Engineering, and Semantic Models}
Experiment tracking records the context needed for replay and auditability~\citep{gamage2025enhancingquantumsoftwaredevelopment}.
Broader quantum software engineering research~\citep{zhao2021quantumsoftwareengineeringlandscapes,piattini2021quantum,9340056}
addresses testing~\citep{9678798} and debugging~\citep{11134329}.
These approaches connect software changes with recorded experimental context and checks
on program behavior. For a cross-provider comparison, the records must also establish
which compiled circuit, placement and outcome representation each measurement describes.
The present paper examines these relationships and identifies checks for failures at
the boundaries between compilation, submission and analysis.

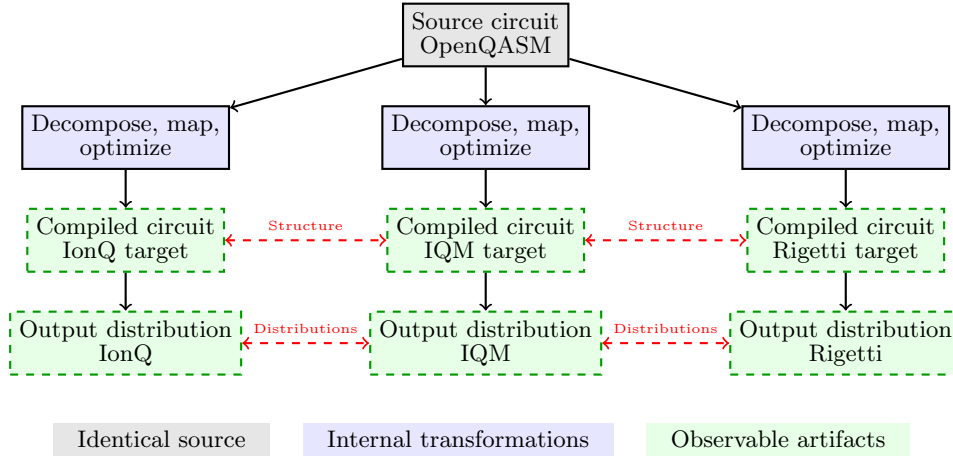
\begin{figure*}[htbp]
\centering
\begin{tikzpicture}[
    node distance=0.5cm and 1.2cm,
    layer/.style={rectangle, draw, thick, minimum width=2cm, minimum height=0.8cm, align=center, font=\scriptsize},
    layerfixed/.style={layer, fill=gray!20},
    layervariable/.style={layer, fill=blue!10},
    layermeasured/.style={layer, fill=green!10, dashed, draw=green!60!black, thick},
    arrow/.style={->, thick},
    comparison/.style={<->, thick, red, dashed}
]
\node[layerfixed] (srcB) {Source circuit\\OpenQASM};
\node[layervariable, below=of srcB] (irB) {Decompose, map,\\optimize};
\node[layermeasured, below=of irB] (trB) {Compiled circuit\\IQM target};
\node[layermeasured, below=of trB] (execB) {Output distribution\\IQM};
\node[layervariable, left=2cm of irB] (irA) {Decompose, map,\\optimize};
\node[layermeasured, below=of irA] (trA) {Compiled circuit\\IonQ target};
\node[layermeasured, below=of trA] (execA) {Output distribution\\IonQ};
\node[layervariable, right=2cm of irB] (irC) {Decompose, map,\\optimize};
\node[layermeasured, below=of irC] (trC) {Compiled circuit\\Rigetti target};
\node[layermeasured, below=of trC] (execC) {Output distribution\\Rigetti};
\draw[arrow] (srcB) -- (irA);
\draw[arrow] (srcB) -- (irB);
\draw[arrow] (srcB) -- (irC);
\draw[arrow] (irA) -- (trA);
\draw[arrow] (irB) -- (trB);
\draw[arrow] (irC) -- (trC);
\draw[arrow] (trA) -- (execA);
\draw[arrow] (trB) -- (execB);
\draw[arrow] (trC) -- (execC);
\draw[comparison] (trA.east) -- (trB.west) node[midway, above, font=\tiny] {Structure};
\draw[comparison] (trB.east) -- (trC.west) node[midway, above, font=\tiny] {Structure};
\draw[comparison] (execA.east) -- (execB.west) node[midway, above, font=\tiny] {Distributions};
\draw[comparison] (execB.east) -- (execC.west) node[midway, above, font=\tiny] {Distributions};
\node[below=0.5cm of execB, font=\scriptsize, align=center] {
    \colorbox{gray!20}{~~Identical source~~} \quad
    \colorbox{blue!10}{~~Internal transformations~~} \quad
    \colorbox{green!10}{~~Observable artifacts~~}
};
\end{tikzpicture}
\caption{From a common source to target-specific realizations. Compilation includes gate
decomposition, qubit mapping and optimization; these steps may be interleaved.
The lower arrows include provider processing and hardware execution, not all of which
is observable. Dashed horizontal arrows identify structural and distributional
comparisons. The available compiled artifacts depend on the submission mode and
provider interface}
\label{fig:layered_model}
\end{figure*}

In classical compiler design, formal semantic models enable reasoning about program
transformations and portability~\citep{DBLP:books/aw/AhoSU86}. The complementary problem
here is to compare observable realizations across multi-provider quantum stacks.
Figure~\ref{fig:layered_model} shows the steps and the two observation points used in this
paper. Identical source circuits pass through target-specific compilation and execution,
but internal transformations are not necessarily visible to the user.

\citet{peltonen2026drift} introduce a CI/CD workflow for detecting
\emph{transpilation drift}: temporal changes in the realization of a circuit for a
\emph{fixed} target caused by toolchain evolution, evaluated at the transpilation layer
only. The provenance framework of \citet{peltonen2026provenance} records the software
and execution context of quantum runs behind a unified multi-provider API.
The present work addresses the complementary problem of \emph{cross-provider realization
divergence}, measured at both the transpilation and execution layers on four commercial
quantum processors. The drift score of \citet{peltonen2026drift} and the divergence
score defined in Section~\ref{framework} form one family of measures. Their references
differ: a temporal baseline versus a concurrent cross-provider artifact.
Section~\ref{sec:implications} discusses a possible extension of that CI/CD gate
from single-target drift detection to portability-aware multi-provider validation.

\textbf{Research focus.} Building on compiler comparison, benchmarking and testing, this
paper evaluates conditional claims from accessible compilation and execution artifacts.
It combines a layered observation model with hardware replications, placement controls
and an examination of provider-interface limits to establish which comparisons those
artifacts support.

\section{Research Methodology}
\label{methodology}

This research follows Design Science Research (DSR)~\citep{10.5555/2017212.2017217,10.2753/MIS0742-1222240302} to design and evaluate a \emph{Divergence Analysis Framework} addressing operational portability in multi-provider quantum workflows. The framework comprises both conceptual elements (a layered observation model and a taxonomy of realization divergence) and an operational toolchain that analyzes observable artifacts from multi-provider executions.

\subsection{Artifact Design}

\Needspace{7\baselineskip}
The toolchain implements a four-stage pipeline:
\begin{enumerate}
    \item \textbf{Transpilation.} Qiskit~\citep{qiskit} compiles source circuits for
    provider-specific gate sets and device topologies. It is widely used for circuit
    manipulation and benchmarking~\citep{nation2025benchmarking}.
    \item \textbf{Structural analysis.} The toolchain parses transpiled
    OpenQASM to extract depth, gate counts
    and qubit mappings. It computes structural divergence pairwise across providers.
    \item \textbf{Execution.} Circuits are submitted to QPUs via the Amazon Braket
    API~\citep{braket2019} with standardized shot counts. Results are retrieved and validated.
    \item \textbf{Divergence quantification.} Measurement distributions are compared,
    and cross-layer correlations link structural transformations to execution outcomes.
\end{enumerate}
The toolchain uses only observable artifacts accessible through provider APIs:
OpenQASM representations, device specifications and measurement results. It requires no
access to internal compiler state or proprietary data. We evaluate its use for
algorithm-specific comparisons and investigation of interface failures. Integration into
CI/CD regression gates is a proposed application, not an evaluated deployment.

\textbf{Submission tracks.} We use two submission modes, called \emph{tracks}.
On the \emph{supported track}, the client compiles to operations accepted by the
provider's interface, leaving further compilation and physical-qubit placement to the
provider. On the \emph{native track}, the client compiles to the device's native gates
and connectivity, and submits the circuit in a verbatim box where available. The box
requests execution without further circuit rewriting, but does not make physical pulses
or every provider-internal operation observable. A supported gate set is therefore an
interface-level operation set, whereas a native gate set describes the device-level
operations targeted by the client.

\subsection{Research Questions}

The framework is demonstrated using representative circuits on commercial quantum platforms and evaluated against three research questions. The first asks what can be claimed about divergence, the second what a provider's interface lets one observe, and the third what goes wrong when heterogeneous components are composed:
\begin{itemize}
    \item \textbf{RQ1:} Which cross-provider realization-divergence claims are portable, and which hold only under stated conditions of metric weighting, placement policy and submission track?
    \item \textbf{RQ2:} What does a provider's interface actually guarantee about an execution, and how does the resulting variation in observability constrain what can be compared across providers?
    \item \textbf{RQ3:} What failure modes arise when heterogeneous compilation and execution components are composed, and what invariants detect them?
\end{itemize}

Pre-registration and the guard discipline described in Section~\ref{evaluation} are the means by which these are answered rather than results in themselves: they are what makes a scored outcome distinguishable from an outcome chosen after the fact.

\textbf{AI-assisted tool usage.}
The authors used GitHub Copilot to assist with code development and review,
test development, reproducibility and artifact-packaging checks, and
manuscript revision. The authors reviewed and validated the resulting
changes and take full responsibility for the analyses, interpretations
and final manuscript.

\section{Divergence Analysis Framework}
\label{framework}

This section presents the Divergence Analysis Framework: its layered observation model, taxonomy of realization divergence, and operational mechanisms for detecting and quantifying realization divergence across quantum software stacks.

\subsection{Design Objectives and Scope}

The framework supports systematic identification and analysis of realization divergence
across heterogeneous platforms. It is designed for quantum software engineers and
algorithm researchers working across providers. Its objectives are to expose transformations,
compare realizations and localize divergence by layer. It relies on accessible artifacts
and records provenance to support reproducibility.

The framework applies to gate-based circuits on cloud-accessible QPUs. Structural
measurements characterize compilation, optimization and mapping; execution measurements
reflect their combined effects with hardware noise, which is not independently identified.

\subsection{Layered Observation Model}

The framework models quantum program execution through four layers of program representation and execution: source language (\(L_{\mathrm{src}}\)), intermediate representations (\(L_{\mathrm{ir}}\)), transpiled target-constrained circuits (\(L_{\mathrm{tr}}\)), and execution (\(L_{\mathrm{exec}}\)):

\[
L = \langle L_{\mathrm{src}}, L_{\mathrm{ir}}, L_{\mathrm{tr}}, L_{\mathrm{exec}} \rangle
\]

The internal layer \(L_{\mathrm{ir}}\) includes transformations such as canonicalization
and basis normalization. These are typically inaccessible in production toolchains.
Our empirical analysis therefore focuses on observable artifacts at \(L_{\mathrm{tr}}\)
and \(L_{\mathrm{exec}}\), the two comparison rows in Figure~\ref{fig:layered_model}.

\textbf{Source language layer (\(L_{\mathrm{src}}\)).}
The source layer captures algorithmic intent in platform-independent form. Here, \(L_{\mathrm{src}}\) is represented using OpenQASM~\citep{cross2017openquantumassemblylanguage}, a vendor-neutral, textual quantum assembly language providing explicit, executable program descriptions over logical qubits, abstract gates, measurements, and control flow. While alternative source representations exist (e.g., Qiskit QuantumCircuit\footnote{Qiskit \texttt{QuantumCircuit} API documentation, \url{https://quantum.cloud.ibm.com/docs/en/api/qiskit/qiskit.circuit.QuantumCircuit}.}, Cirq\footnote{Cirq project documentation, \url{https://quantumai.google/cirq}.}, Pytket\footnote{Pytket distribution page, \url{https://pypi.org/project/pytket/}.}), OpenQASM was selected for its role as a standardized interchange format with broad toolchain support and human-inspectable syntax. \(L_{\mathrm{src}}\) serves as the \textbf{reference realization}: cross-platform comparisons require identical \(L_{\mathrm{src}}\) artifacts.

\textbf{Transpiled target-constrained layer (\(L_{\mathrm{tr}}\)).}
\(L_{\mathrm{tr}}\) represents device-specific lowering: target gate set conversion, qubit layout selection, connectivity-constrained routing, and often scheduling to device clocks. This is the "compiled circuit" in practice and the primary locus for cross-platform structural divergence.

Key divergence mechanisms at \(L_{\mathrm{tr}}\) include approximate synthesis, heuristic mapping/routing (altering depth, gate count, idle times), scheduling policies (affecting decoherence), and alternative native bases (yielding distinct physical behavior under noise).

\textbf{Execution layer (\(L_{\mathrm{exec}}\)).}
The execution layer captures observed measurement distributions. These depend on upstream
transformations, device state and readout behavior. Measuring this layer separately from
\(L_{\mathrm{tr}}\) does not by itself attribute a difference to hardware or software.

\textbf{Divergence reference models.}
Divergence is measured relative to explicit references. \textbf{Source-referenced analysis}
compares structural metrics to the source circuit and output distributions to its ideal
simulation. \textbf{Pairwise analysis} compares provider realizations directly. Neither
structural differences nor differences in noisy outputs prove a failure of ideal equivalence.

\subsection{Realization Divergence Taxonomy}

Table~\ref{tab:semdiv} presents the framework's taxonomy of observable divergence types. Rather than assuming equivalence, it explicitly captures how provider-specific transformations alter program realization across layers. Transpilation-layer divergence arises from representations, decompositions, and optimizations altering circuit structure; execution-layer divergence captures sampling behavior and hardware state sensitivity. This layered taxonomy enables systematic detection and attribution using concrete artifacts.

\begin{table*}
\centering
\caption{Realization Divergence Taxonomy: Measured Divergence Types}
\label{tab:semdiv}
\footnotesize
\begin{tabularx}{\textwidth}{l >{\raggedright\arraybackslash}p{2.6cm} X X X}
\toprule
Layer & Divergence Type & Description & Observable Artifacts & Engineering Relevance \\
\midrule
\multirow{5}{*}{\raisebox{2.2em}[0pt][0pt]{\textbf{Transpilation (\(L_{\mathrm{tr}}\))}}}
    & Structural Circuit Divergence & Different gate structures & Transpiled circuits & Changed resource requirements \\
  & Logical-to-Physical Mapping & Divergent qubit layout selection & Qubit mapping metadata & Increased noise sensitivity \\
    & Gate Decomposition & Different native gate set decompositions & Native gate sequences & Different physical error exposure \\
    & Optimization-Induced Variation & Transpiler optimizations altering circuit structure & Depth/gate count changes & Sensitivity to compiler configuration \\
  & Circuit-Dependent Scaling	& Divergence magnitude varies by algorithm structure & Cross-circuit structural metrics	& Algorithm-specific portability boundaries \\
\midrule
\multirow{2}{*}{\raisebox{0.5em}[0pt][0pt]{\textbf{Execution (\(L_{\mathrm{exec}}\))}}}
  & Distributional Divergence & Differences in output probability distributions & Measurement distributions & Statistical inconsistency, loss of signal \\
  & Calibration Sensitivity & Results depend on transient hardware state & Measurement distributions, temporal variance & Reduced reproducibility \\
\bottomrule
\end{tabularx}
\end{table*}

\subsection{Operationalization and Quantification}
The framework analyzes artifacts at two primary observation points: transpiled circuits (\(L_{\text{tr}}\)) produced by client-side Qiskit compilation and execution results (\(L_{\text{exec}}\)) retrieved via Amazon Braket API—both accessible using standard SDK artifacts without requiring access to provider-internal compilation stages or proprietary toolchain data. The source representation (\(L_{\mathrm{src}}\)) serves as the reference realization for measuring provider-induced divergence.

We adapt the baseline-relative structural drift score of \citet{peltonen2026drift}
to source-relative and cross-provider comparisons. Structural divergence between a
target realization $A$ and a reference realization $R$ is quantified as the weighted mean of relative metric
differences:
\[
D_{\text{struct}}(A, R)
= \sum_{k=1}^{K} w_k \,
\frac{\left| m_k^{(A)} - m_k^{(R)} \right|}{m_k^{(R)}},
\qquad \sum_k w_k = 1,
\]
over $K{=}3$ structural metrics: circuit depth, two-qubit gate count, and
total gate count (size). Unlike the workshop formulation, which floors the denominator
at one, we exclude metrics with zero-valued reference and renormalize the remaining
weights. This preserves the interpretation as an average relative difference, which
is undefined for a zero reference. Consequently, changes from zero to a positive count
are not reflected in the aggregate score and must be assessed from the individual
metrics. The retained metrics receive equal weights.
$D_{\text{struct}} = 0$ indicates equality of the retained metrics, not circuit identity;
values are interpreted as average absolute relative differences (e.g.,
$D_{\text{struct}} = 1.0$ corresponds to a 100\% average difference
relative to the reference). The score is reference-directed: when $R$ is
the source circuit $L_{\mathrm{src}}$, it captures both expansion and contraction without
retaining their direction. For pairwise comparison, the realization with smaller depth
is the reference; ties are broken by size and then two-qubit count. This deterministic
convention makes the pairwise score symmetric but does not make it a mathematical distance.

Execution-layer detection compares output probability distributions using Jensen–Shannon (JS) divergence~\citep{1207388}:
\[
D_{\text{dist}}(p_1, p_2) = \text{JS}(P_{p_1}, P_{p_2})
\]
With finite shot counts, many outcomes go unobserved. Kullback--Leibler divergence is
infinite when an outcome has positive probability in its first argument but zero in its
reference distribution. JS divergence is symmetric and remains finite and bounded in
this case~\citep{1207388}. Comparing output distributions is standard practice in quantum
benchmarking, although the measure varies between
studies~\citep{Lubinski_2023,tomesh2022supermarqscalablequantumbenchmark}.
Alternatives include Hellinger distance and total variation distance. JS provides an
information-theoretic interpretation of distributional difference~\citep{1207388}.
For algorithms with well-defined ideal outputs, such as Bernstein–Vazirani, we compare
against noiseless simulation. For variational or stochastic algorithms, we use pairwise
comparison.

\subsection{Metric Validity Conditions}

Each metric is reported with the regime in which it carries information, and the
conditions below are preconditions on individual claims rather than a one-time filter
applied to the dataset.

\textbf{Reference and uncertainty.} JS uses base-two logarithms and lies in $[0,1]$.
Against a fixed point mass at the target, it depends only on the target probability $p$:
\[
\operatorname{JS}(P,\delta_{\mathrm{target}})
= h_2\!\left(\frac{1+p}{2}\right)-\frac{h_2(p)}{2},
\]
where $h_2$ is binary entropy. This is a monotone decreasing function of $p$.
Reporting $P(\mathrm{target})$ makes success probability and its chance baseline explicit;
it does not add information or restore statistical power lost by JS. Proximity to unity
alone does not invalidate an ordering. We assess uncertainty using per-run intervals.
Comparisons against diffuse references or between two sampled distributions have different
finite-sampling behavior and are not interpreted using a point-mass comparison's uncertainty.

\textbf{Composite structural scores.} $D_{\text{struct}}$ combines three per-circuit
quantities --- depth, gate count and two-qubit gate count --- into one number, and doing so
requires deciding how much each contributes. That decision is the \emph{metric weighting},
and this study weights the three equally. The choice is a free parameter rather than a
measurement. We therefore re-score every result over a grid of alternative weightings
and compare them with equal weighting. The grid contains all non-negative weights
summing to one in steps of $0.1$, giving 66 candidate metrics. Across this grid, the
BV8--BV12 structural inversion under lowest-index placement remains present on Emerald
and Cepheus and absent on Garnet. The full circuit ranking changes with the weights.
This check concerns that specific within-device comparison, not cross-device clustering.
We report the tested ordering with its conditions; a magnitude or full ranking requires
the weighting and reference to be stated alongside it.

\textbf{Statistical power.} Shot counts are chosen for each experimental comparison
according to the effect size it must resolve. For threshold tests, we compute the range of true values that
the planned shot count cannot distinguish from the threshold. We report this undecidable
range as part of the design. An experiment that cannot resolve the effect its question requires
must be redesigned before execution.

\section{Framework Demonstration}
\label{demonstration}

This section demonstrates how the Divergence Analysis Framework connects observations
of compiled structure with execution results on commercial quantum platforms. The Bell
example illustrates why the observation layer matters; the BV case study introduces the
comparisons tested in the evaluation.

\subsection{Experimental Setup}

We describe the platforms, compilation controls and circuit selection used in the
demonstration. These establish the common workflow against which provider-specific
realizations are compared; Section~\ref{evaluation} then evaluates claims through a
separate pre-registered experimental campaign: a planned set of hardware runs across
circuits, devices, submission tracks and repetitions.

\subsubsection{Quantum Platforms and Toolchain}

We demonstrate the framework using the operational toolchain described in Section~\ref{methodology}, applying it across four commercial quantum platforms accessed via Amazon Braket.
All experiments follow a controlled cross-provider workflow starting from identical OpenQASM circuits as the semantic reference (\(L_{\mathrm{src}}\)). Circuits are transpiled using Qiskit v2.5.1 (optimization level 3) targeting provider-recommended supported gate sets and device topologies retrieved from Amazon Braket backend specifications; the campaign harness refuses to run under any other version. We fix seeds for stochastic passes and explicitly enforce hardware-specific coupling maps during transpilation, and preliminary experiments confirmed that omitting topology constraints artificially inflates structural divergence. Seed fixing alone does not deliver reproducibility, so we state the narrower property we can support. Repeated transpilation is bit-identical on the native track under a forced layout in the tested cases. On the supported track, QAOA depth varied by twelve percent across repeated runs. Under compiler-default placement, different embeddings can carry identical depth, size and two-qubit counts --- so agreement on those counts is not evidence that the same circuit was produced. Reproducibility claims here therefore name the track and placement policy as well as the seed.

Hardware execution covers IonQ Forte-1 (trapped-ion, all-to-all connectivity), IQM Emerald
and Garnet (superconducting, lattice connectivity), and Rigetti Cepheus (superconducting,
lattice connectivity). Emerald and Garnet share a native gate set but differ in device
size, available connectivity and calibration state; they are not identical-topology
controls. Archived Rigetti Ankaa-3 profiles additionally enter compilation-only analyses.
Common Amazon Braket orchestration standardizes collection, but does not control provider
scheduling, calibration or internal compilation. For each run we retain the available
circuit, mapping, backend and measurement artifacts.

\subsubsection{Test Circuit Selection}

Circuits are selected to expose realization divergence from qubit connectivity, native
gate sets, routing constraints and compilation strategies. Selection follows recurring
\emph{compilation stressors} in the literature~\citep{Lubinski_2023, Quetschlich_2023}.
Greenberger–Horne–Zeilinger (GHZ) states test global entanglement, and Quantum Fourier
Transform (QFT) circuits test dense long-range interactions. BV
circuits provide oracle-based constructions with deterministic outputs. The Quantum
Approximate Optimization Algorithm (QAOA) supplies structured graph circuits.
These patterns expose differences in connectivity, gate-set expressivity and routing
overhead across architectures~\citep{li2019tacklingqubitmappingproblem,wack2021qualityspeedscalekey}.

Circuits are evaluated at two scales: small instances (4–6 qubits) isolating gate-set expressivity, and medium instances (8–12 qubits) introducing non-trivial routing on constrained topologies. Larger circuits are excluded to ensure all experiments execute on real hardware under comparable conditions. Table~\ref{tab:testcircuits} summarizes the selected circuits.

\begin{table}[t]
\centering
\caption{Circuits used in the framework demonstration of Section~\ref{demonstration}. The
pre-registered hardware campaign of Section~\ref{evaluation} uses the wider set in
Table~\ref{tab:campaigncircuits}.}
\label{tab:testcircuits}
\footnotesize
\begin{tabular}{p{1.3cm}cp{2.5cm}p{2.5cm}}
\toprule
\textbf{Circuit} & \textbf{Qubits} & \textbf{Category} & \textbf{Reference} \\
\midrule
BV & 8, 12 & Oracle-based & Deterministic output \\
GHZ & 4, 6 & Global entanglement & Ideal GHZ distribution \\
QAOA (p=1, ring) & 8 & Variational & MaxCut value, ideal statevector \\
QFT & 10 & Quantum transform & Uniform distribution \\
\bottomrule
\end{tabular}
\end{table}

\subsection{Bell Circuit: Layer-Specific Interpretation}
\label{sec:bell_interpretation}

We revisit the motivating example from Section~\ref{sec:bell_motivation} using the
framework's observation layers. Figure~\ref{fig:bell_compilation} shows the native-gate
realizations. At the supported gate set --- the layer most users program against, and the layer at
which portability is normally claimed --- all four evaluated devices return the Bell
circuit unchanged, at the same two gates and the same depth. Nothing diverges, and a study
that measured only this layer would report perfect portability.

The divergence appears one layer down. Native compilation produces six single-qubit
rotations around one entangler on IQM Emerald. On Rigetti Cepheus it produces seven
rotations around one entangler drawn from a different basis.
The two-qubit structure eliminates routing, so this is gate-set expressivity alone. The
third device, IonQ Forte-1, is absent for a different reason, and the reason is the
interesting part. Its native gates are not in the client-side compiler's standard library,
so a basis named from the vendor's own device description cannot be resolved by the
vendor-neutral toolchain at all. They are reachable, but only through a vendor-supplied
plugin that defines the gates as bespoke objects, which is a different compilation path
from the one every other device in this study uses. Native-layer portability for IonQ
Forte-1 therefore requires leaving the common toolchain. A study that compiles through
only one interface may record this gap as missing data rather than as a finding.

Source-level portability therefore does not imply operational portability, and the layer at
which a portability claim is made determines whether it can be false.

\subsection{Representative Case Study: Bernstein–Vazirani Algorithm}

We illustrate cross-layer analysis using the 12-qubit Bernstein–Vazirani instance (BV12).
BV has deterministic ideal behavior, enabling direct divergence
measurement~\citep{SUN2026101108}. Its shallow logical depth limits noise-dominated effects.
It is also sensitive to qubit mapping and routing, with sufficient scale to induce
non-trivial overhead on constrained topologies. BV12 stresses compilation pipelines while
remaining executable on all evaluated platforms, allowing joint analysis of transpilation
and execution layers.

\subsubsection{Transpilation-Layer Divergence}

The stored native BV12 artifacts have (depth, two-qubit gate count, total gate count)
values of $(45,18,83)$ on Emerald, $(61,35,132)$ on Garnet and $(73,29,129)$ on
Cepheus~\citep{peltonen2026divartifact}.
These are observations of particular compiled artifacts, not hardware-quality scores.
They do not support general clustering by vendor or architecture: under equal weighting,
Garnet is structurally closer to Cepheus than to Emerald. IonQ has no comparable native
artifact in this workflow. A supported-track artifact cannot fill that gap because it
precedes different provider-side transformations.

\subsubsection{Execution-Layer Divergence}

For BV12, the ideal reference is a point mass on the hidden string. Both JS divergence
and target probability summarize departure from that reference; their monotone relationship
is given in Section~\ref{framework}. Neither alone identifies the cause of a departure.
The following evaluation uses per-run confidence intervals and comparisons across circuit
sizes to test the placement-based explanation, with 1000 shots in the main campaign and
8000 in the placement-matched follow-up experiment.

\section{Evaluation}
\label{evaluation}

This section uses a hardware campaign and compilation-only follow-up experiments to
address the research questions from Section~\ref{methodology}. The preceding demonstration
illustrates the observations; the experiments below test claims about their interpretation.

\subsection{Evaluation Setup}
\label{sec:evaluation_setup}

The evaluation separates three tasks: testing circuit orderings (RQ1), inspecting what
provider interfaces expose (RQ2), and identifying failures between toolchain stages (RQ3).
The main hardware campaign supplies execution records for all three tasks. Compilation-only
placement experiments and a hardware follow-up then test explanations of the RQ1 results.

\textbf{Circuit selection.} Whereas the demonstration illustrates individual cases, the
campaign varies circuit size within a family to distinguish a size-based explanation
from an embedding-based one. Table~\ref{tab:campaigncircuits} lists the wider circuit set.
Under the tested lowest-index placement on Emerald, embedding radius and circuit size
predict different orderings for BV8 versus BV10 and BV12: the smaller circuit can require
longer connections. The GHZ sweep provides a contrast in which the two accounts predict
the same direction. BV scale labels count measured data bits; each source circuit also
contains an ancilla used by the oracle.

\begin{table}[t]
\centering
\caption{Main hardware campaign set. Each circuit was submitted to Emerald, Garnet and
Cepheus on supported and native tracks at two repetitions. IonQ was evaluated separately
on a smaller supported-track control set. Bell provides a minimal-circuit reference.}
\label{tab:campaigncircuits}
\footnotesize
\begin{tabular}{p{1.6cm}p{3.2cm}p{2.4cm}}
\toprule
\textbf{Family} & \textbf{Scales (qubits)} & \textbf{Role} \\
\midrule
BV & 5, 8, 10, 12, 16 & Discriminating sweep \\
GHZ & 4, 6, 15 & Contrast sweep \\
QFT & 10 & Single instance \\
QAOA ($p{=}1$, ring) & 8 & Single instance \\
Bell & 2 & Noise floor \\
\bottomrule
\end{tabular}
\end{table}

\textbf{Hardware comparisons.} The main campaign comprises 11 circuits, three devices,
two submission tracks and two repetitions: $11\times3\times2\times2=132$ runs, plus
18 supported-track noise-floor runs. A placement-matched follow-up adds 12 Cepheus runs
to compare submission tracks after replacing the original fixed native placement with
topology-aware client placement. It covers three BV circuits, two tracks and two
repetitions. A separate all-to-all control on IonQ tests whether the Emerald ordering
also appears when direct connectivity removes the need for routing. It comprises one
Bell probe and eight supported-track runs: BV8, BV10, BV12 and GHZ6 at two repetitions.
The four devices therefore have different experimental roles and coverage.

\textbf{Placement experiment.} The compilation-only experiment asks whether the
radius--overhead association persists when we deliberately change the qubit layout.
In a BV oracle, the ancilla interacts with selected data qubits, forming a star-shaped
interaction graph. Its realized radius is the largest shortest-path distance from the
mapped ancilla to a participating data qubit on the device coupling graph. We compute
it from the final layout reported by the compiler. Two-qubit overhead is the compiled
two-qubit gate count minus the source count.

We compare three placement procedures: a fixed lowest-index graph traversal, the
compiler's default placement, and randomized paths through connected qubits without
revisiting a qubit. The randomized procedure supplies two draws, labelled Random A and
Random B in the figure (\texttt{random-0} and \texttt{random-1} in the data); these are
not different algorithms. Three pseudorandom seeds initialize both the stochastic
compiler choices and the generation of the random layouts. The design crosses five BV
sizes, four device profiles (including archived Ankaa-3), three seeds and four placement
variants. It yields 239 recorded embeddings out of 240 planned combinations. These
shared circuits and profiles make the correlations descriptive, not independent hardware
replications. Exact seeds, the missing combination and generation settings are listed
in \texttt{REPLICATION.md} in the artifact~\citep{peltonen2026divartifact}.

The 239-embedding radius analysis uses Qiskit only. A separate placement-policy
comparison uses both Qiskit and tket~\citep{Sivarajah_2020} to test whether the
BV8--BV12 structural inversion depends on placement policy in both compilers.
This comparison does not extend the two-compiler coverage to the hardware campaign
or the radius analysis.

\textbf{Uncertainty and rank analysis.} We estimate per-run shot uncertainty by drawing
new count samples from the observed outcome frequencies (multinomial bootstrap).
Unless stated otherwise, 1000 resamples give the 2.5th and 97.5th percentiles, forming a
95\% confidence interval. These intervals describe sampling uncertainty within a run,
not changes in device calibration over time. Replicated orderings require disjoint
individual intervals in each repetition.

The separate descriptive rank analysis compares two-qubit overhead with mean execution
JS across the two repetitions. Its exact permutation tests are one-sided and exploratory,
without multiplicity adjustment. Thus this summary of average ranks serves a different
purpose from the per-repetition ordering tests.

\textbf{Chance-level criterion for rank analysis.} To restrict the descriptive BV rank
analysis to cases with evidence of above-random correct-answer probability, we retain a
circuit--device--track combination only when its lower 95\% bootstrap confidence bound
exceeds the uniform-random baseline $2^{-b}$ in both repetitions. Here $b$ is the number
of measured data bits, so $2^{-b}$ is the probability of obtaining the single correct
bit string under uniform random sampling. We use 400 resamples for this criterion.
This is an inclusion rule for that analysis, not a validity requirement for JS divergence;
the resulting correlations describe only the retained combinations. Failure to meet the
criterion does not establish random behavior or the absence of meaningful differences.

\textbf{Submission-track ratios.} The placement-matched comparison uses the log-delta
interval specified in Amendment 9c, a dated addition to the registered analysis plan
archived in \path{campaigns/phase-b/predictions.md}~\citep{peltonen2026divartifact}.
This interval quantifies uncertainty in the ratio of correct-answer probabilities.
For independent
target proportions $p_s,p_n$ from $N_s,N_n$ shots and ratio $r=p_s/p_n$, its limits are
\[
\exp\!\left(\ln r \pm 1.96
\sqrt{\frac{1-p_s}{N_s p_s}+\frac{1-p_n}{N_n p_n}}\right).
\]
Here $s$ and $n$ denote supported and native submission. Recomputation from the raw
counts reproduces all six recorded intervals at their printed precision. An independent
check using 100000 binomial resamples per track and percentile limits agrees at two
decimal places. This checks sensitivity to the interval method rather than replacing
the registered method. Resampling seeds are recorded in the artifact's replication guide.

\textbf{Registration chronology.} The original predictions preceded the first BV
submissions. Amendment 1 revised predictions after affected submissions but, according
to its contemporaneous record, before the affected results were inspected. Amendment 2
introduced scoring categories after Garnet and Cepheus repetition-one results were
collected. Later replication and placement-matched comparison rules were fixed before the corresponding
follow-up data. We distinguish those prospective tests from exploratory analyses and
post-submission amendments; not every rule was pre-registered before the first run.
The dated predictions, amendments and outcomes are retained in
\path{campaigns/phase-b/predictions.md} and \path{docs/ablation_grid.md}
in the artifact~\citep{peltonen2026divartifact}.

\subsection{RQ1: What Can Be Claimed, and Under What Conditions}

\textit{Which cross-provider realization-divergence claims are portable, and which hold only under stated conditions of metric weighting, placement policy and submission track?}

We first establish which structural differences are visible and whether they agree
with execution rankings. We then test an embedding-based explanation in four steps:
vary placement to examine radius and overhead, change the placement policy to test the
structural ordering, check the predicted orderings on hardware, and reassess submission
tracks after improving native placement. This sequence separates descriptive
associations from targeted tests of their interpretation.

\textbf{Observation layer and metric sensitivity.}
All available native-track artifacts differ from their source on the selected structural
metrics. Supported-track artifacts often agree on those metrics. Such agreement is a
valid observation at that layer, but it cannot establish identical native realizations
or execution behavior.
The run records behind every value in this section, and the scripts that derive them, are in the research artifact~\citep{peltonen2026divartifact}.

Table~\ref{tab:presence} reports source-referenced structural scores, not pairwise
device distances. Bell and GHZ have zero supported-track scores in the available cells;
BV and QFT have nonzero source-referenced scores even where their compiled metrics agree
across devices. QAOA also changes on the supported track. These observations distinguish
change from the source from differences between providers, which are different questions.

\input{presence_table}

On the all-to-all device, basis conversion and optimization can change circuit structure
even without routing. The supported artifacts alone do
not isolate which transformation produced a particular change.

The BV12 example does not support grouping devices by architecture. Across all 66 tested
weightings, the IQM pair is never the closest native pair for that circuit. The robust
ordering established by the sensitivity analysis is instead the within-device BV8--BV12
inversion under lowest-index placement, discussed below.

\textbf{Descriptive agreement with execution.} We compare ranks within each circuit
family and device, applying the chance-level criterion to BV target probabilities.

Table~\ref{tab:rankagreement} compares two-qubit overhead with mean execution JS within
each family and device. The BV chance gate removes Cepheus BV12 and BV16 on the native
track; the reported BV correlations are conditional on this selection.
Five rows have perfect rank agreement; Emerald BV has $\rho=0.70$.
Only Garnet BV has an unadjusted exact $p<0.05$, and the six tests are exploratory.
At three circuits the minimum attainable one-sided $p$ is $1/6$, so a perfect rank
correlation alone is not strong inferential evidence. Averaging the repetitions here is
descriptive; it is not used to establish the independently replicated orderings below.

\input{rank_table}

Cepheus illustrates why uncertainty must be assessed for the actual comparison. Native
BV8 and BV12 have JS estimates of 0.9629 and 0.9853 in repetition one, and 0.9595 and
0.9663 in repetition two. Their individual bootstrap intervals overlap in both repetitions;
the predicted BV8-greater-than-BV12 ordering is not established by this criterion.
The point estimates run in the opposite direction. We do not dismiss them merely because
JS is near one, and do not generalize this unresolved comparison to every BV pair.
The original saturation-based scoring rationale in Amendment 2 is retained in the
registration history described in Section~\ref{sec:evaluation_setup}; the present
comparison instead uses the per-run intervals.

These descriptive comparisons motivate a narrower question: can the way a circuit is
embedded in the device explain why a smaller instance sometimes diverges more than a
larger one? The following experiments test that explanation.

\subsubsection{Assigned-Placement Experiment}
\label{sec:mechanism}

This experiment tests whether the radius--overhead association persists across deliberately
varied placements of the same circuits on the same device profiles.
Across 60 earlier compilation configurations (four BV sizes, three devices and five seeds), radius
correlates with two-qubit overhead at Spearman $0.903$, against $0.336$ for circuit size.
These are a separate dataset from the 239 assigned-policy embeddings. The latter experiment
tests whether the association persists when placement is deliberately varied.
Over the 239 recorded embeddings described above, the corresponding coefficients are
$0.805$ and $0.492$. The policy with the smallest mean radius also has the smallest mean
overhead, and the association is positive within each policy separately.
Figure~\ref{fig:radius_overhead} reports these results. Assignment intervenes on placement,
not radius alone: other properties of the embedding change with it.
The association therefore extends beyond one fixed placement procedure, making realized
embedding a useful descriptor of compilation cost in this dataset. The next experiment
tests whether a particular size ordering also survives a change of placement policy.

\input{radius_scatter}

\subsubsection{Placement-Policy Test}

A registered comparison tested whether the BV8--BV12 structural inversion --- greater
structural divergence for the smaller circuit --- depended on placement policy.
In the tested comparison, it appears under lowest-index
placement on Emerald and Cepheus, but not Garnet, and disappears under compiler-default
placement. Qiskit and tket agree on which device--placement combinations exhibit this
inversion~\citep{peltonen2026divartifact}. The 66-weight analysis
preserves this specific ordering, not the complete ranking of circuit magnitudes.
Thus the inversion is a property of the tested circuit--device--placement combination,
not circuit size alone. The hardware experiment tests whether the predicted execution
ordering also recurs across repetitions.

\subsubsection{Hardware Replication and Control Devices}

The hardware test asks whether the size orderings predicted under lowest-index placement
recur in execution, and whether a different embedding produces a different ordering.
Under lowest-index placement, the predicted Emerald BV8--BV12 ordering has disjoint
individual bootstrap intervals in both repetitions. Garnet shows the opposite ordering,
also with disjoint intervals in both repetitions. The rule requiring independent
replication was fixed after repetition one and before repetition two, not before all
hardware observations. Table~\ref{tab:ordering} reports the differences in JS against
the ideal distribution at 1000 shots per run. Garnet's available qubits changed between
repetitions, so a stable ordering does not imply stable divergence levels.

\begin{table}[t]
\centering
\caption{Execution-layer ordering on two devices from the same vendor, with the same native
gate set, circuits and submission path, under lowest-index placement. Values are
$\operatorname{JS}(\mathrm{BV8})-\operatorname{JS}(\mathrm{comparison})$.
Positive means the eight-qubit instance diverged more.}
\label{tab:ordering}
\footnotesize
\begin{tabular}{llcc}
\toprule
Device & Comparison & Rep 1 & Rep 2 \\
\midrule
\multirow{2}{*}{IQM Emerald} & BV8 vs BV12 & $+0.0443$\textsuperscript{a} & $+0.0600$\textsuperscript{a} \\
                          & BV8 vs BV10 & $+0.0727$\textsuperscript{a} & $+0.0827$\textsuperscript{a} \\
\midrule
\multirow{2}{*}{IQM Garnet} & BV8 vs BV12 & $-0.1788$\textsuperscript{a} & $-0.0694$\textsuperscript{a} \\
                          & BV8 vs BV10 & $-0.0303$ & $-0.0192$ \\
\bottomrule
\end{tabular}

\vspace{2pt}
{\footnotesize \textsuperscript{a} individual bootstrap 95\% confidence intervals disjoint.
Unmarked comparisons are direction-consistent but unresolved by this criterion.}
\end{table}

The reversal rules out an explanation based only on source size or source gate count.
It supports the embedding-based account, but does not exclude device-specific noise,
qubit quality or interactions between those factors and placement. The compiled gate
counts themselves differ across devices and are part of the proposed explanation.
On the all-to-all IonQ control, radius is one by construction and the Emerald inversion
is absent in both repetitions. This limited supported-track control does not establish
monotonicity across all circuit sizes or isolate connectivity from hardware technology.
The supported conclusion is a replicated, device-dependent ordering under the stated
placement policy: BV8 diverges more than BV12 on Emerald and less on Garnet. The
opposite directions are the evidence against circuit size as a sufficient explanation.

\subsubsection{Matched-Placement Comparison of Submission Tracks}

This follow-up asks whether the earlier supported-track advantage persists when native
submission uses a topology-aware placement rather than the original fixed layout.
The original native track used a fixed lowest-index graph traversal, while the supported
track delegated placement to the provider. Thus the track comparison also changed the
placement policy. On Cepheus, giving Qiskit its own default layout matches the vendor's
two-qubit count on three of five circuits and reduces it on the other two. Equal counts
do not imply equal native programs or equal physical error exposure.

The follow-up compares topology-aware, calibration-blind client placement with provider
compilation on BV5, BV8 and BV10 at 8000 shots per run. ``Matched'' refers to removing
the original unfavorable placement policy, not identical physical qubits or pulse programs.
Table~\ref{tab:matched} reports supported-to-native target-probability ratios. All six
upper confidence limits are below the registered factor-of-two threshold. The registered
label ``second-order'' describes the factor of two category, not negligible practical impact:
BV5 favors the supported track by $1.66\times$ and $1.71\times$.

\input{matched_table}

Native-track advantages observed for BV8 and BV10 in repetition one did not both clear
unity independently in repetition two. The data therefore establish neither uniform
vendor superiority nor uniform native-track superiority. They bound the observed track
advantage in this comparison, without identifying calibration-aware selection as its sole cause.

\textbf{Answer to RQ1.} Three findings survive the tested conditions. First, realized
radius is positively associated with two-qubit overhead across the placement variants.
Second, the BV8--BV12 structural inversion survives all 66 tested weightings on Emerald
and Cepheus under lowest-index placement, but disappears under compiler-default placement.
On hardware, Emerald and Garnet reproduce opposite BV8--BV12 execution orderings under
lowest-index placement. Third, the placement-matched Cepheus comparison finds no uniform
winner between submission tracks, while all six upper ratio limits remain below two.
These results support specific conditional comparisons: robustness to metric weighting
does not imply robustness to placement, and a structural ordering does not automatically
predict execution on another device. A transferable report must therefore specify the
observation layer, reference, placement, weighting and uncertainty with its result.

\subsection{RQ2: What a Provider's Interface Guarantees}

\textit{What does a provider's interface actually guarantee about an execution, and how does the resulting variation in observability constrain what can be compared across providers?}

What can be measured about an execution is a provider variable rather than a constant, and
the levels were established by exercising them. IQM and Rigetti return the compiled program
alongside a per-shot record, permitting comparison of the submitted and returned programs
and checks on verbatim handling, though not independent verification of physical pulses.
In the evaluated interface, IonQ returns outcome probabilities with no per-shot record
and no compiled program, preventing that structural comparison. A study that assumes a
uniform observability level will silently compare a measured quantity on one device with an
unavailable one on another.

A published native gate set does not determine what a device will execute. Rigetti
advertises \texttt{rx}, \texttt{rz} and \texttt{cz} without qualification but, inside a
verbatim box, accepts \texttt{rx} only at $\pm\pi/2$ and $\pm\pi$; IQM's parametrised
rotation accepts an arbitrary angle. A circuit can therefore be native by gate name and
rejected in practice, which is why the native track in this study carries a basis restricted
by measured behaviour rather than by the published list. Three further constraints surfaced
only at the submission boundary: IQM indexes qubits from one where the client-side compiler
indexes densely from zero, IonQ enforces a minimum shot count, and the collected timing
records do not resolve queueing from execution, preventing that latency attribution here.

The strongest case is a contract that stayed still while the machine moved.
Table~\ref{tab:generation} compares Rigetti Ankaa-3, retired during the study, with its
successor Rigetti Cepheus. Their supported-operation lists are identical; the native
entangler changed from \texttt{iswap} to \texttt{cz}. The supported-operation list alone
does not reveal that change. Native-layer two-qubit counts have Ankaa-to-Cepheus ratios
from $0.63$ to $1.20$ across seven circuits, with no uniform direction. These count
changes do not by themselves imply changes in execution quality. They motivate measuring
below the supported-operation abstraction.

\input{generation_table}

\textbf{Answer to RQ2.} The interface guarantees less than it appears to, and in three
distinct ways: it under-specifies what will execute, it under-reports what did, and it can
remain constant across a change in the machine beneath it. Each bounds what a cross-provider
comparison may claim. Returned compiled programs are available on IQM and Rigetti
but absent from the evaluated IonQ interface. A native-basis claim must state admissible parameter values
alongside the gate names. A portability result must also name the device generation
measured, because the advertised contract may not record that the generation changed.

\subsection{RQ3: Failure Modes of a Composed Toolchain}

\textit{What failure modes arise when heterogeneous compilation and execution components are composed, and what invariants detect them?}

A multi-provider measurement pipeline composes a client-side compiler, a vendor service, a
device and an analysis stage, each with its own contract. Across this study the failures at
those joins fell into three modes, distinguished by what the interface did with an
instruction rather than by which component was at fault.

\emph{Needed but undeclared.} A constraint governs execution but appears nowhere in the
device description, so a valid-looking program is refused --- the restricted rotation
angles above are the clearest instance. This mode is benign in the sense that it announces
itself: the submission fails.

\emph{Declared but unconsumed.} A constraint is published and the client does not read it.
IQM Emerald publishes weekday execution windows; a batch submitted outside one waited hours
while an otherwise identical batch inside one returned in seconds, which looks like queue
variance and is a schedule. This mode also announces itself, but as a false signal rather
than an error, and it is attributed to the device unless the published metadata is
consulted.

\emph{Accepted but overridden.} An instruction is accepted by one stage and silently
discarded by a later one. A forced qubit placement was applied by one compiler pass and
then re-placed by a subsequent pass that begins with a placement step of its own. The
forced-layout and compiler-default branches then produced identical results, invalidating
that intended policy contrast. Each pass behaved as documented --- the fault is in the composition, not in
either pass --- which is what makes the mode general rather than a property of one library.
Inspecting an output circuit alone did not expose the defect: it remained valid,
in-basis and connectivity-respecting. The violated relationship was between the requested
placement and its realization, exposed by the policy comparison.

The same signature appears at the analysis boundary. IonQ returning probabilities where the
collector expected shot counts yields an empty distribution recorded as a successful run; a
distribution compared against a reference of a different bitstring width gives two disjoint
supports and a maximal divergence that looks like a measurement. In each case the artifact
is well formed and the relationship is wrong.

The reporting pipeline provided another instance. A findings entry cited an analysis file
that did not contain the reported value, although both artifacts were individually well
formed. We therefore extended the boundary checks to reporting: values regenerated from
run records are compared against the findings ledger in the test suite. The comparison
specifies the finding, device, track, repetition and column, rather than hard-coding the
expected value. This is an instance of the RQ3 failure mode within our own measurement
pipeline, not a separate contribution about scientific writing.

The corresponding checks have different evidence requirements. Published constraints can
be checked before submission: qubit identifiers, permitted shot counts and execution
windows. An undocumented angle restriction instead requires a capability probe or an
observed rejection, followed by a recorded constraint; rereading the same metadata cannot
discover an omitted rule. Placement checks compare requested and realized mappings and
test contrasts expected to remain distinct under the experiment's contract. Arbitrary
input changes need not change an optimized output.

At the analysis boundary, count totals must match the recorded shots. Probability-only
responses require separate normalization and shot-count handling. Comparisons require
consistent outcome width and bit ordering, not identical observed support: two valid
samples can have disjoint observed outcomes. The implemented width check catches the
reported width mismatch, but does not verify every aspect of outcome semantics.

\textbf{Answer to RQ3.} The composed toolchain exhibits both explicit rejections and silent relational failures.
The practical response is to validate the specific contract at each boundary, while
recognizing that unobserved failure modes remain outside those checks.

\section{Discussion}
\label{sec:discussion}

\subsection{Why a Divergence Magnitude Is Not a Device Property}

A structural score characterizes a circuit, compiler configuration, reference and device
profile jointly. It is not an intrinsic property of a device. In the weighting experiment,
only 12 of 66 native-track weightings reproduce the equal-weight circuit ranking; the
minimum Spearman correlation with that ranking is zero. This shows sensitivity to the
chosen aggregation, not that the underlying metric values lack meaning.

Placement and device generation introduce separate dependencies. Changing placement
alters routing costs, while changing the native entangler can alter synthesis costs even
when the supported-operation list is unchanged (Table~\ref{tab:generation}). A magnitude
is interpretable when those conditions are specified, but cannot be transferred to an
unstated configuration. The specific BV inversion survives the tested weightings under
lowest-index placement; that result does not imply that orderings generally survive
changes of placement, device or generation.

\subsection{Interpretation and Scope}

The framework makes aspects of operational portability observable through structural
metrics and output distributions. These measurements do not establish quantum-state
fidelity or causal attribution to particular noise sources. A zero score in
Table~\ref{tab:presence} denotes agreement on selected metrics, not proof that the
complete circuits or their physical realizations are identical.

The placement intervention strengthens the structural explanation by deliberately varying
embeddings. It does not independently manipulate radius while holding every other
embedding property fixed. The hardware reversal and all-to-all control are consistent
with that explanation, but qubit quality, hardware technology and provider transformations
remain alternative or interacting causes. The placement-matched comparison likewise bounds an observed
track difference; it is not a causal estimate of calibration-aware placement alone.

\section{Implications for Quantum Software Engineering}
\label{sec:implications}

The discussion above establishes how to interpret the measurements. This section draws
out their consequences for engineering practice and software quality; it introduces no
additional experimental results.

\subsection{Implications for Practice}

Practitioners should record the observation layer, source and compiled artifacts, reference
distribution, metric weights, physical layout, device profile and software versions with
each comparison. A seed alone is insufficient. Capability checks should cover native
parameter domains and returned result types, not just gate names.

An earlier comparison of Qiskit 0.45.2 and 2.3.0, conducted before the campaign was
pinned to 2.5.1, found measurable structural differences between transpiler versions.
This illustrates why software provenance belongs alongside the circuit and device
profile in a reproducible comparison.

These records can support regression investigation and portability assessment. They also
provide a basis for developing automated CI gates, extending the earlier drift
workflow~\citep{peltonen2026drift}, although a deployed multi-provider CI workflow remains
to be evaluated. The Cepheus comparison illustrates how submission choices can be assessed
within a device. Extending that assessment to provider selection requires comparable
workloads, explicit placement and compilation conditions, and uncertainty estimates across
the candidate providers. In practice, these conditions should accompany each comparison
so that its relevance to a deployment decision can be assessed.

\subsection{Implications for Software Quality}

The findings above are stated in the vocabulary of quantum compilation. They are more
usefully read as statements about established quality attributes, and Table~\ref{tab:quality}
maps them onto five.

\begin{table}[t]
\centering
\caption{Engineering quality concerns informed by the study and the limits of that evidence}
\label{tab:quality}
\footnotesize
\begin{tabularx}{\linewidth}{>{\raggedright\arraybackslash}p{2.5cm} X}
\toprule
\textbf{Attribute} & \textbf{What this study contributes, and its condition} \\
\midrule
Portability &
Record both structural and execution observations. Equality at a supported abstraction
does not establish equality at the native or execution layer (Table~\ref{tab:presence}). \\
Interoperability &
A provider's published interface is a partial specification of what it will execute.
A native gate set can name a rotation without constraining its angle domain.
A supported-operation list remained identical across a generation that replaced the
native entangler (Table~\ref{tab:generation}). Execution records also vary by provider, making the submitted-versus-executed
comparison available on some devices and impossible on others.
Composing such interfaces admits three failure modes --- a contract
needed but undeclared, declared but unconsumed, or accepted and then overridden --- of
which only the first produces an immediate rejection in the observed examples.
Metadata checks and capability probes provide complementary evidence; an interoperability
claim covers the interfaces and parameter values actually exercised
(Section~\ref{evaluation}). \\
Reproducibility &
What must be pinned for bit-identical transpilation is now stated rather than assumed: a
fixed seed is insufficient, and the placement policy has to be fixed alongside it.
Native forced-layout output is bit-identical in the tested cases; supported-track QAOA
metrics vary between invocations. The same seed under compiler-default placement returns
different embeddings with identical depth,
size and two-qubit counts (Section~\ref{demonstration}). \\
Analysability &
Returned compiled programs permit additional checks on IQM and Rigetti, but are absent
from the evaluated IonQ interface. The placement-matched comparison bounds one performance trade-off
associated with verbatim submission on Cepheus (Table~\ref{tab:matched}); it does not
establish that auditability is free. \\
CI reliability &
Structural metrics are candidates for cross-provider regression signals, extending
single-target drift detection~\citep{peltonen2026drift}. A gate would need a specified
track, reference, placement policy and tolerance; its operational effectiveness remains
to be evaluated. \\
\bottomrule
\end{tabularx}
\end{table}

These quality concerns are interdependent. Submission choices affect both the evidence
available for analysis and the transformations applied before execution. The Cepheus
comparison illustrates why analysability and execution performance should be assessed
together: topology-aware native placement provides greater control over the submitted
circuit, but neither submission track has a consistent correct-answer advantage across
the tested cases. Quality assessment therefore requires explicit consideration of both
observability and measured outcomes.

The interface failures also shift attention from individual components to their
composition. A circuit may compile and a submission may succeed while a placement
requirement is ignored or an outcome representation is misinterpreted. Quality assurance
should therefore check relationships between source circuits, compiled artifacts,
submitted programs and returned results, alongside checks on each artifact separately.

The common shape is that each attribute becomes measurable only once its validity condition
is stated. A portability figure without its layer, a reproducibility claim without its
placement policy, or a regression gate without its submission track are all quantities that
can be computed but cannot reliably support comparisons beyond their recorded conditions.

\section{Limitations and Threats to Validity}
\label{sec:threats}

The findings are bounded by the devices, circuits and toolchains exercised in this study.
We first identify threats to the reported measurements, then delimit their scalability
and applicability beyond the evaluated NISQ setting.

\subsection{Threats to the Empirical Findings}

\textbf{Circuit and execution coverage.} Limited circuit coverage and changes in device
state could make an observed ordering specific to a workload or execution session.
The main campaign and placement-matched follow-up therefore use two repetitions; the
separate IonQ Bell probe has one. Headline orderings require disjoint intervals separately
in both repetitions, without pooling shots to obtain narrower intervals. Only the
descriptive rank table averages execution levels. These checks support recurrence of the
specified orderings, not stable divergence levels: Garnet's available qubits and execution
levels changed between repetitions. The results remain limited to the tested workloads
and do not characterize longer-term drift.

\textbf{Transpiler and toolchain dependencies.} Compiler versions, stochastic choices,
placement and metric weighting can change structural comparisons. The campaign harness
enforces Qiskit v2.5.1, and repeated compilation checks establish bit-identical native
output under forced layouts in the tested cases. Placement and weighting experiments
test sensitivity beyond that fixed configuration; Qiskit and tket agree on the tested
placement-policy partition. These measures establish repeatability and specific
sensitivity results, not general compiler independence. Supported-track QAOA metrics vary
between invocations, and compiler-default placement can change embeddings without changing
depth, size or two-qubit counts. Vendor-internal transformations remain partly unobserved;
matching gate counts does not isolate synthesis quality. The 239-placement dataset also
includes archived Ankaa-3 profiles and one missing combination. Its shared circuits,
profiles and seeds make its correlations descriptive, not independent hardware replications.

\textbf{Uneven coverage of the all-to-all device.} IonQ contributes nine execution records
against fifty or more for each other device, limiting cross-device comparisons. Its
per-shot cost constrained coverage to a targeted control: whether the Emerald ordering
also appears where the star radius cannot vary. We report missing observations as absent
cells rather than measured zeros (Table~\ref{tab:presence}) and restrict interpretation
to the circuits executed. This reporting prevents missing evidence from being mistaken
for agreement, but does not remedy the coverage imbalance. The control informs the
specified ordering comparison, not a balanced provider ranking or a universal
circuit-class claim; connectivity also remains confounded with hardware technology.

\textbf{Track coverage and confounding.} The supported and native tracks differ in
placement and provider processing, so performance differences cannot be attributed to
submission mode alone. The Cepheus follow-up addresses the original fixed-placement
disadvantage by replacing lowest-index placement with topology-aware, calibration-blind
native placement. This tests whether the track difference persists after that change,
but does not match physical qubits, pulse programs or calibration conditions. Its ratios
therefore characterize the tested submission choices, not the isolated effect of
calibration-aware selection. Coverage remains incomplete: IonQ has no comparable native
artifacts in the common toolchain and no BV16 executions on either track. Native
observations also remain specific to their recorded device generation and profile.

\textbf{Interface failures.} Silent failures can produce plausible measurements from
inconsistent artifacts, threatening the reported structural and execution comparisons.
Checks on count totals, outcome widths and placement contracts address distinct failure
modes, while existing tests cover identified cases and selected ledger-to-artifact
comparisons. These checks reduce known measurement risks and connect selected reported
values to their evidence. They do not validate every scientific claim or detect every
possible interface failure; successful analysis and passing tests remain evidence of
the relationships checked, not completeness of the measurement pipeline.

\textbf{Registration and researcher degrees of freedom.} Post-submission and
post-observation amendments create opportunities for choices influenced by observed
results. We preserve the original predictions and amendment chronology
(Section~\ref{sec:evaluation_setup}), label the rank analyses and post-audit comparisons
as exploratory, and assess later repetitions using rules fixed before those repetitions.
This separates prospective tests from exploratory interpretation and tests whether selected
orderings recur in new executions. It cannot make earlier observations prospective or
eliminate selection effects; replicated findings retain the scope of the comparisons
selected for follow-up.

\textbf{Statistical interpretation.} Finite shots and analysis selection limit the
precision and scope of execution comparisons. We report per-run intervals, make the BV
rank-inclusion rule explicit, and label permutation tests as exploratory and unadjusted
for multiple comparisons. A binomial-bootstrap sensitivity check agrees with the
registered log-delta ratio intervals at the reported precision. These measures quantify
conditional sampling uncertainty and check sensitivity to the interval method; they do
not cover unobserved calibration changes. Rank correlations apply only to the retained
cases. Overlapping individual intervals do not establish equivalence, and ratios below
the registered factor-of-two threshold do not establish negligible differences.
Near-unity JS values remain informative through their monotone relation to target
probability, so uncertainty is assessed for the comparison rather than inferred from
proximity to the metric's bound.

\subsection{Scalability and Generalization Beyond NISQ}

The evaluation covers the circuit sizes and NISQ-era devices reported above. It does not
establish the scalability of artifact collection and analysis for substantially larger
circuits, nor validate the observed orderings on fault-tolerant hardware.

The layered model suggests a possible extension to logical-to-fault-tolerant compilation.
Such an extension would need to account for code-specific decomposition, logical-qubit
placement, and scheduling against decoding and magic-state resources. Candidate structural
measures include T-count, logical depth and logical-qubit overhead. These would require
their own reference models, weighting analysis and empirical validation. Reduced logical
error rates would also change the operating regime of distributional metrics; they would
not by themselves establish cross-provider comparability or causal attribution.
Applicability beyond NISQ therefore remains an open validation question, not a result
of this study.

\section{Future Work}
\label{sec:future_work}

The limitations in Section~\ref{sec:threats} motivate five directions for further work:
\begin{itemize}
    \item \textbf{Cross-toolchain validation.} Repeat the placement and metric-weighting
    analyses across compiler versions, optimization settings and native-gate interfaces.
    This would test whether the reported validity conditions hold beyond the pinned toolchain.
    \item \textbf{Physical attribution.} Incorporate calibration data, pulse-level
    representations and noise models to distinguish compilation effects from specific
    physical mechanisms. The present distributional comparisons do not establish that attribution.
    \item \textbf{Coverage and scalability.} Extend the circuit families and all-to-all
    device coverage, and measure collection and analysis costs at larger scales.
    Dynamic circuits would test applicability beyond the evaluated workloads.
    \item \textbf{Longitudinal replication.} Repeat campaigns across calibration cycles
    and hardware generations. This would test the stability of orderings beyond two
    repetitions and distinguish temporal drift from cross-provider differences.
    \item \textbf{Fault-tolerant compilation.} Extend the layered observation model to
    logical-to-fault-tolerant compilation, accounting for error-correction schemes,
    logical-qubit placement and resource-constrained scheduling. Develop and validate
    reference models and structural measures such as T-count, logical depth and
    logical-qubit overhead, and reassess distributional metrics under logical-error
    regimes. This would test which validity conditions transfer beyond the NISQ setting.
\end{itemize}

\section{Conclusion}
\label{conclusion}
This paper measured realization divergence across heterogeneous quantum platforms and asked which of the resulting claims can be carried anywhere else. By modeling divergence across language, transpilation, and execution layers, and implementing an integrated toolchain to extract, compare, and quantify divergence using observable artifacts, the work makes the conditions attached to such claims explicit rather than implicit.

Grounded in Design Science Research, the framework and toolchain make the three
contributions in Section~\ref{introduction} testable through observable artifacts.
The evaluation follows common circuit specifications through compilation and execution,
connecting the observed differences to their measurement conditions.

For RQ1, the Emerald BV8--BV12 execution ordering and its opposite on Garnet replicate
under lowest-index placement. The placement experiment supports an embedding-dependent
explanation without identifying individual noise mechanisms. The placement-matched Cepheus comparison
keeps all supported-to-native target-probability ratios below the registered factor of
two, but neither establishes equivalence nor isolates calibration-aware selection.
Structural weighting sensitivity supports specific conditional orderings, not general
architecture clustering or a universal device ranking.

For RQ2, provider interfaces expose different evidence and omit constraints relevant to
execution. A supported-operation list can remain unchanged while the native entangler
changes. Comparisons must therefore name the observed artifact, parameter domain and
device generation rather than assuming uniform visibility.

For RQ3, composed toolchains exhibit both explicit rejections and silent inconsistencies
between otherwise valid artifacts. Metadata validation, capability probes, layout checks
and outcome-encoding checks address different parts of that problem. Their value is in
testing explicit relationships, not in guaranteeing that every failure mode is covered.

Operational portability must therefore be measured at stated observation layers and under
explicit validity conditions. This paper provides evidence for which realization-divergence
claims remain defensible and which interface
checks are needed to make them reproducible.

\section*{Artifact Availability}
The framework implementation, experimental data, and analysis scripts are provided in the accompanying research artifact~\citep{peltonen2026divartifact} to support reproducibility and community extension.
The repository contains raw run records, archived device profiles, analysis outputs and
regeneration scripts. Tests compare selected findings-ledger values against analysis
artifacts and reject distribution-width mismatches. These checks do not cover every
claim or establish byte-identical regeneration of every artifact. Dated registrations
and findings ledgers are preserved as historical records; the manuscript distinguishes
their original scoring rationale from the statistical interpretation adopted here.

\section*{Acknowledgments}
We thank QMill Oy for providing the Amazon Braket access under which the hardware
campaigns reported here were executed. Funding is stated separately below.

\section*{Statements and Declarations}

\textbf{Funding.} This work was supported by Business Finland through project EM4QS
(155/31/2024), and by the Research Council of Finland through the Finnish Quantum Flagship
(359240, JYU) and project DEQSE (349945).

\textbf{Competing interests.} The first author is employed by QMill Oy, which provided the
Amazon Braket access under which the hardware campaigns reported here were executed. QMill
Oy neither manufactures nor operates quantum hardware and is not among the providers
evaluated in this study. None of the evaluated providers had any role in the design of the
study, in the execution or analysis of the experiments, or in the decision to publish. The
remaining authors declare no competing interests.

\textbf{Data availability.} The measurement records, device profiles and derived analysis
files supporting the findings of this study, including every pre-registered clause and its
scored outcome, are available in the accompanying research
artifact~\citep{peltonen2026divartifact}. No data were obtained from third parties under
restriction; the raw execution results were produced by the authors on commercially
available quantum hardware.

\textbf{Code availability.} The analysis toolchain, campaign harness and scripts
for generated figures and tables are included in the same research
artifact~\citep{peltonen2026divartifact}.

\textbf{Author contributions.} \textbf{Jouni Peltonen}: conceptualization, methodology,
software, validation, formal analysis, investigation, data curation, visualization, writing
--- original draft. \textbf{Vlad Stirbu}: supervision, writing --- review and editing.
\textbf{Tommi Mikkonen}: supervision, writing --- review and editing. All authors read and
approved the final manuscript.

\bibliography{references}

\end{document}

%% file: presence_table.tex
\begin{table}[t]
\centering
\caption{Observed structural differences from the source. Each cell reports whether the structural score against the source circuit is exactly zero (unshaded) or non-zero (shaded); magnitudes are not quoted because the composite score is weighting-dependent. Zero denotes agreement on depth, size and two-qubit count, not circuit identity or an execution guarantee. All available native entries differ from their source on these metrics. A dash is an absent artifact: the all-to-all device has no native-track entry because its native basis does not resolve on the vendor-neutral toolchain, and it entered the study in a separate campaign covering fewer circuits.}
\label{tab:presence}
\scriptsize
\setlength{\tabcolsep}{3pt}
\begin{tabular}{lcccccccc}
\toprule
 & \multicolumn{4}{c}{Supported gate set} & \multicolumn{4}{c}{Native gate set} \\
\cmidrule(lr){2-5}\cmidrule(lr){6-9}
Circuit & Emerald & Garnet & Cepheus & Forte-1 & Emerald & Garnet & Cepheus & Forte-1 \\
\midrule
Bell (2) & \cellcolor{white}$0$ & \cellcolor{white}$0$ & \cellcolor{white}$0$ & \cellcolor{white}$0$ & \cellcolor{black!12}$>0$ & \cellcolor{black!12}$>0$ & \cellcolor{black!12}$>0$ & \cellcolor{white}-- \\
GHZ (4) & \cellcolor{white}$0$ & \cellcolor{white}$0$ & \cellcolor{white}$0$ & \cellcolor{white}-- & \cellcolor{black!12}$>0$ & \cellcolor{black!12}$>0$ & \cellcolor{black!12}$>0$ & \cellcolor{white}-- \\
GHZ (6) & \cellcolor{white}$0$ & \cellcolor{white}$0$ & \cellcolor{white}$0$ & \cellcolor{white}$0$ & \cellcolor{black!12}$>0$ & \cellcolor{black!12}$>0$ & \cellcolor{black!12}$>0$ & \cellcolor{white}-- \\
GHZ (15) & \cellcolor{white}$0$ & \cellcolor{white}$0$ & \cellcolor{white}$0$ & \cellcolor{white}-- & \cellcolor{black!12}$>0$ & \cellcolor{black!12}$>0$ & \cellcolor{black!12}$>0$ & \cellcolor{white}-- \\
BV (5) & \cellcolor{black!12}$>0$ & \cellcolor{black!12}$>0$ & \cellcolor{black!12}$>0$ & \cellcolor{white}-- & \cellcolor{black!12}$>0$ & \cellcolor{black!12}$>0$ & \cellcolor{black!12}$>0$ & \cellcolor{white}-- \\
BV (8) & \cellcolor{black!12}$>0$ & \cellcolor{black!12}$>0$ & \cellcolor{black!12}$>0$ & \cellcolor{black!12}$>0$ & \cellcolor{black!12}$>0$ & \cellcolor{black!12}$>0$ & \cellcolor{black!12}$>0$ & \cellcolor{white}-- \\
BV (10) & \cellcolor{black!12}$>0$ & \cellcolor{black!12}$>0$ & \cellcolor{black!12}$>0$ & \cellcolor{black!12}$>0$ & \cellcolor{black!12}$>0$ & \cellcolor{black!12}$>0$ & \cellcolor{black!12}$>0$ & \cellcolor{white}-- \\
BV (12) & \cellcolor{black!12}$>0$ & \cellcolor{black!12}$>0$ & \cellcolor{black!12}$>0$ & \cellcolor{black!12}$>0$ & \cellcolor{black!12}$>0$ & \cellcolor{black!12}$>0$ & \cellcolor{black!12}$>0$ & \cellcolor{white}-- \\
BV (16) & \cellcolor{black!12}$>0$ & \cellcolor{black!12}$>0$ & \cellcolor{black!12}$>0$ & \cellcolor{white}-- & \cellcolor{black!12}$>0$ & \cellcolor{black!12}$>0$ & \cellcolor{black!12}$>0$ & \cellcolor{white}-- \\
QFT (10) & \cellcolor{black!12}$>0$ & \cellcolor{black!12}$>0$ & \cellcolor{black!12}$>0$ & \cellcolor{white}-- & \cellcolor{black!12}$>0$ & \cellcolor{black!12}$>0$ & \cellcolor{black!12}$>0$ & \cellcolor{white}-- \\
QAOA (8, $p{=}1$) & \cellcolor{black!12}$>0$ & \cellcolor{black!12}$>0$ & \cellcolor{black!12}$>0$ & \cellcolor{white}-- & \cellcolor{black!12}$>0$ & \cellcolor{black!12}$>0$ & \cellcolor{black!12}$>0$ & \cellcolor{white}-- \\
\bottomrule
\end{tabular}
\end{table}

%% file: rank_table.tex
\begin{table}[t]
\centering
\caption{Agreement between structural and execution rank, within circuit family and device: two-qubit overhead versus mean JS across two repetitions. Each family uses the same ideal-reference construction, with circuit-dependent width. BV cells failing the chance gate in either repetition are excluded. The one-sided exact permutation $p$-values are exploratory and unadjusted for multiple comparisons. Their minimum is $1/n!$, so $\rho=+1.00$ at $n=3$ carries $p=0.167$, insufficient to reject the permutation null at 0.05.}
\label{tab:rankagreement}
\footnotesize
\begin{tabular}{llccc}
\toprule
Family & Device & Circuits $n$ & Rank correlation $\rho$ & Exact $p$ \\
\midrule
BV & IQM Emerald & 5 & $+0.70$ & $0.117$ \\
BV & IQM Garnet & 5 & $+1.00$ & $0.008$ \\
BV & Rigetti Cepheus & 3 & $+1.00$ & $0.167$ \\
GHZ & IQM Emerald & 3 & $+1.00$ & $0.167$ \\
GHZ & IQM Garnet & 3 & $+1.00$ & $0.167$ \\
GHZ & Rigetti Cepheus & 3 & $+1.00$ & $0.167$ \\
\bottomrule
\end{tabular}
\end{table}

%% file: radius_scatter.tex
\begin{figure}[t]
\centering
\begin{tikzpicture}
\begin{groupplot}[
    group style={group size=2 by 2, horizontal sep=0.7cm, vertical sep=1.0cm,
                 xlabels at=edge bottom, ylabels at=edge left},
    width=0.48\linewidth, height=4.3cm,
    tick label style={font=\scriptsize}, title style={font=\small},
    label style={font=\small},
    xlabel={Realised star radius},
    ylabel={Two-qubit gate overhead},
    xmin=0.5, xmax=8.5, ymin=-2, ymax=62,
    xtick={1,2,3,4,5,6,7,8}, ytick={0,20,40,60},
    grid=major, grid style={gray!25},
]
\nextgroupplot[title={Compiler default}]
\addplot[only marks, mark=*, mark size=1.5pt,
         color=blue!70!black]
    coordinates {(1,0) (1,0) (1,0) (1,0) (1,0) (1,0) (1,0) (1,0) (1,0) (1,3) (1,3) (1,3) (2,3) (2,3) (2,1) (2,3) (2,3) (2,3) (2,3) (2,3) (2,3) (2,4) (2,8) (2,8) (2,3) (2,3) (2,3) (2,3) (2,3) (2,3) (2,3) (2,3) (2,1) (2,8) (2,8) (2,8) (3,6) (2,3) (3,6) (3,9) (2,6) (3,6) (3,6) (2,6) (2,6) (3,13) (4,12) (2,13) (3,6) (3,6) (4,9) (3,9) (3,6) (4,7) (3,9) (3,9) (2,4) (3,17) (4,9) (2,16)};
\node[anchor=north west, font=\scriptsize, align=left, fill=white,
      inner sep=2pt] at (rel axis cs:0.03,0.97)
    {$\bar{r}=2.1$, $\bar{o}=5.0$\\
$\rho=+0.77$, $N=60$};
\nextgroupplot[title={Random A}]
\addplot[only marks, mark=square*, mark size=1.5pt,
         color=orange!80!black]
    coordinates {(1,6) (1,9) (1,9) (2,10) (2,12) (2,4) (1,6) (1,9) (1,9) (2,9) (1,12) (1,15) (2,15) (3,16) (4,14) (3,18) (4,14) (4,16) (3,21) (3,18) (4,33) (3,22) (3,22) (3,17) (3,22) (3,21) (4,13) (4,22) (3,17) (4,17) (3,24) (3,16) (4,35) (4,25) (2,20) (2,17) (3,16) (5,28) (3,14) (4,14) (5,20) (5,19) (5,17) (3,19) (8,43) (4,33) (4,26) (4,28) (6,35) (5,23) (5,25) (5,17) (5,17) (5,26) (5,28) (4,30) (8,42) (5,35) (5,44) (5,33)};
\node[anchor=north west, font=\scriptsize, align=left, fill=white,
      inner sep=2pt] at (rel axis cs:0.03,0.97)
    {$\bar{r}=3.5$, $\bar{o}=20.3$\\
$\rho=+0.72$, $N=60$};
\nextgroupplot[title={Random B}]
\addplot[only marks, mark=triangle*, mark size=1.5pt,
         color=green!55!black]
    coordinates {(2,12) (2,12) (1,9) (1,9) (1,9) (1,6) (2,13) (1,12) (1,9) (1,6) (1,6) (2,9) (6,32) (3,13) (4,18) (3,18) (2,16) (4,14) (5,26) (3,18) (2,9) (3,20) (3,17) (4,19) (5,34) (3,18) (3,18) (4,24) (3,15) (5,17) (4,32) (4,19) (2,15) (2,16) (2,19) (2,10) (8,36) (4,23) (3,25) (5,23) (3,18) (6,21) (6,40) (4,27) (4,18) (5,23) (3,27) (6,27) (8,40) (6,31) (6,37) (5,17) (5,26) (8,56) (6,36) (4,42) (7,35) (6,32) (5,33)};
\node[anchor=north west, font=\scriptsize, align=left, fill=white,
      inner sep=2pt] at (rel axis cs:0.03,0.97)
    {$\bar{r}=3.7$, $\bar{o}=21.4$\\
$\rho=+0.87$, $N=59$};
\nextgroupplot[title={Lowest index}]
\addplot[only marks, mark=diamond*, mark size=1.5pt,
         color=red!70!black]
    coordinates {(2,12) (1,12) (1,12) (1,9) (1,9) (1,9) (2,13) (2,13) (2,13) (2,14) (2,14) (2,14) (5,27) (5,27) (4,23) (3,15) (3,15) (2,15) (7,35) (6,37) (6,37) (6,31) (5,31) (5,27) (3,19) (4,19) (3,17) (3,18) (3,18) (4,18) (5,26) (4,28) (4,26) (4,15) (3,15) (3,15) (5,11) (5,11) (4,11) (4,28) (3,28) (3,28) (7,22) (5,24) (5,24) (5,23) (5,23) (5,23) (3,34) (3,30) (4,34) (6,29) (4,27) (5,29) (5,29) (5,29) (5,29) (6,35) (6,31) (6,43)};
\node[anchor=north west, font=\scriptsize, align=left, fill=white,
      inner sep=2pt] at (rel axis cs:0.03,0.97)
    {$\bar{r}=3.9$, $\bar{o}=22.2$\\
$\rho=+0.71$, $N=60$};
\end{groupplot}
\end{tikzpicture}
\caption{Realised star radius against two-qubit gate overhead over 239 Qiskit embeddings of five BV sizes on four device profiles, including archived Ankaa-3. Each point is one circuit--device--seed--placement configuration; coincident points overlap. Panels share axis scales. Random A and B are two draws from the same randomized-path procedure, not different algorithms. $\bar{r}$ and $\bar{o}$ denote mean radius and overhead, $\rho$ is the within-panel Spearman correlation, and $N$ counts configurations. These are compilation results. Across all embeddings, the radius--overhead correlation is $+0.805$ against $+0.492$ for circuit size versus overhead. Placement assignment changes multiple embedding properties; it does not isolate radius alone}
\label{fig:radius_overhead}
\end{figure}
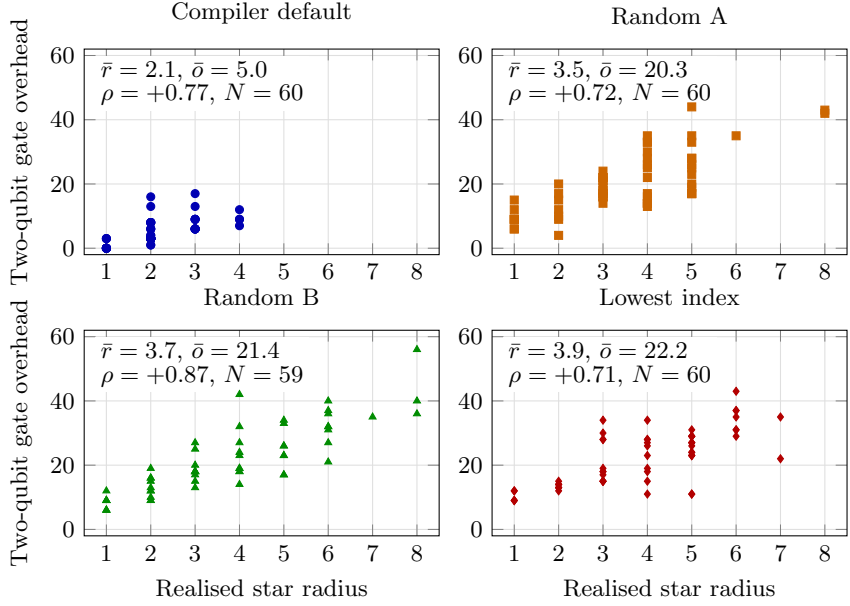

%% file: matched_table.tex
\begin{table}[t]
\centering
\caption{Matched-placement comparison of submission tracks. The supported track delegates placement to the provider; the matched native track uses the client-side compiler's topology-aware but calibration-blind layout. Physical qubits and native programs are not identical across tracks. The ratio is supported-track target probability over matched-native target probability on Rigetti Cepheus, so a value above one favours the supported track. All six log-delta 95\% upper confidence limits are below the registered $2\times$ threshold. This bounds the observed track advantage; it does not establish equivalence or isolate calibration effects from other provider transformations.}
\label{tab:matched}
\footnotesize
\begin{tabular}{llccc}
\toprule
Circuit & Rep & $P_{\mathrm{supported}}$ & $P_{\mathrm{native}}$ & Ratio (95\% CI) \\
\midrule
BV (5) & 1 & $0.6299$ & $0.3785$ & $1.66\times$ [1.61, 1.72] \\
BV (5) & 2 & $0.6795$ & $0.3967$ & $1.71\times$ [1.66, 1.77] \\
BV (8) & 1 & $0.0589$ & $0.0905$ & $0.65\times$ [0.58, 0.73] \\
BV (8) & 2 & $0.0900$ & $0.0984$ & $0.91\times$ [0.83, 1.01] \\
BV (10) & 1 & $0.0520$ & $0.0714$ & $0.73\times$ [0.64, 0.82] \\
BV (10) & 2 & $0.0795$ & $0.0684$ & $1.16\times$ [1.04, 1.30] \\
\midrule
\multicolumn{5}{l}{\footnotesize Registered threshold: $2\times$. Highest observed: $1.71\times$.} \\
\bottomrule
\end{tabular}
\end{table}

%% file: generation_table.tex
\begin{table}[t]
\centering
\caption{A stable contract over a changed device. Rigetti Ankaa-3 and its successor Rigetti Cepheus publish identical supported-operation lists, while the native entangler changed from \texttt{iswap} to \texttt{cz}. Two-qubit gate counts are measured at the native gate set. Values are arithmetic means over three compiler seeds; individual compiled circuits have integer gate counts. The ratio has no uniform direction. The unchanged supported-operation list does not expose these native-layer differences. Counts alone do not establish execution quality. Ankaa-3 was retired during the study and cannot be re-measured.}
\label{tab:generation}
\footnotesize
\begin{tabular}{lccc}
\toprule
 & \multicolumn{2}{c}{Mean two-qubit gate count} & \\
\cmidrule(lr){2-3}
Circuit & Ankaa-3 (\texttt{iswap}) & Cepheus (\texttt{cz}) & Ankaa/Cepheus \\
\midrule
GHZ (6) & $7.3$ & $9.7$ & $0.76$ \\
BV (5) & $17.0$ & $16.0$ & $1.06$ \\
BV (8) & $34.7$ & $41.3$ & $0.84$ \\
BV (10) & $20.0$ & $31.7$ & $0.63$ \\
BV (12) & $30.0$ & $30.3$ & $0.99$ \\
BV (16) & $44.3$ & $37.0$ & $1.20$ \\
QFT (10) & $140.7$ & $131.0$ & $1.07$ \\
\bottomrule
\end{tabular}
\end{table}